\documentclass[a4paper,final,conference]{IEEEtran}

\usepackage[utf8]{inputenc}
\usepackage{graphicx}
\usepackage{amssymb}
\usepackage[cmex10]{amsmath}
\usepackage{color}
\usepackage{theorem}
\usepackage{cite}
\usepackage{url}
\usepackage{breakurl}
\usepackage[english]{babel}
\usepackage[english = american]{csquotes}
\usepackage{mathtools}
\usepackage[ruled,vlined,titlenumbered]{algorithm2e}

\usepackage[margin=0.67in]{geometry}

\makeatletter
\newcommand{\removelatexerror}{\let\@latex@error\@gobble}
\makeatother

\IEEEoverridecommandlockouts
\MakeOuterQuote{"}

\definecolor{institut_color_orig}{rgb}{0.8,0.3,0.4}
\definecolor{blue_aw}{rgb}{0.3,0.3,1.0}

\usepackage{pgf}
\usepackage{tikz}
\usetikzlibrary{matrix,chains,positioning,arrows,calc,decorations,plotmarks,patterns, fit,backgrounds}
\usepackage{pgfplots}
\usetikzlibrary{external}                       
\usepackage{tkz-graph}
\pgfplotsset{compat=1.3}
\tikzstyle{help lines}=[black!20,dashed]

\pgfplotscreateplotcyclelist{mylist}{red,blue,black,yellow,brown}

\pgfdeclarepatternformonly{my north east lines}{\pgfqpoint{-1pt}{-1pt}}{\pgfqpoint{8pt}{8pt}}{\pgfqpoint{6pt}{6pt}}%
{
  \pgfsetlinewidth{0.4pt}
  \pgfpathmoveto{\pgfqpoint{0pt}{0pt}}
  \pgfpathlineto{\pgfqpoint{6.1pt}{6.1pt}}
  \pgfusepath{stroke}
}
\definecolor{light_gray}{rgb}{0.6,0.6,0.6}
\definecolor{awgray}{rgb}{0.7,0.7,0.7}
\definecolor{awgray_dark}{rgb} {0.4,0.4,0.4}

\tikzset{
    >=stealth',
    mycircle/.style={circle, draw=gray, very thick, text width=.1em, minimum height=1.5em, text centered},
    mycircle_small/.style={circle,draw=awgray_dark,fill = awgray_dark, inner sep=0,minimum size=.6em},
    mycircle_small_black/.style={circle,draw=black,fill = black, inner sep=0,minimum size=.6em},
    mybox/.style={rectangle,rounded corners,draw=black, thick,text width=1em,minimum height=4em,minimum width=4em,text centered},
    mybox_small/.style={rectangle,rounded corners,draw=black, thick,text width=1em,minimum height=2em,minimum width=2em,text centered},
    mybox_vec/.style={rectangle,rounded corners,draw=black, thick,text width=1em,minimum height=0.7em, minimum width=4em,text centered},
    mybox_vec_short/.style={rectangle,rounded corners,draw=black, thick,text width=1em,minimum height=0.7em, minimum width=2em,text centered},
    pil/.style={->, thick, shorten <=2pt, shorten >=2pt,},
}

\newtheorem{theorem}{Theorem}

\newtheorem{lemma}{Lemma}

\newtheorem{construction}{Construction}

 \newcommand{\qed}{\hfill \mbox{\raggedright \rule{.07in}{.1in}}}

\newcommand{\Fqm}{\ensuremath{\mathbb F_{q^m}}}

\newcommand{\Fq}{\ensuremath{\mathbb F_{q}}}

\newcommand{\F}{\ensuremath{\mathbb F}}

\newcommand{\OCompl}[1]{\ensuremath{\mathcal{O}({#1})}}

\SetAlgoCaptionSeparator{.} 

\DeclareMathOperator{\rk}{rk}

\renewcommand{\vec}[1]{\ensuremath{\mathbf{#1}}}
\newcommand{\Mat}[1]{\ensuremath{\mathbf{#1}}}

\newcommand{\A}{\Mat{A}}
\newcommand{\B}{\Mat{B}}
\newcommand{\C}{\Mat{C}}

\newcommand{\I}{\mathbf I}

\newcommand{\0}{\vec{0}}

\newcommand{\mycode}[1]{\ensuremath{\mathcal{#1}}}

\newcommand{\codelinearArb}[1]{\ensuremath{[#1]}}

\newcommand{\codelinearRank}[1]{\ensuremath{[#1]_q^\fontmetric{R}}}

\newcommand{\MRDlinq}[1]{\ensuremath{\mycode{MRD}[#1]_q}}

\newcommand{\fontmetric}[1]{\mathsf{#1}}

\newcommand{\Subspacedist}[1]{d_s(#1)}

\newcommand{\myspace}[1]{\mathcal{#1}}

\newcommand{\Grassm}[1]{\myspace{G}_q(#1)}

\newcommand{\quadbinom}[2]{\ensuremath{
{#1
\brack
#2}_q
}}

\newcommand{\quadbinoms}[2]{\ensuremath{
{#1
\brack
#2}_{q_s}
}}

\begin{document}

\title{Vector Network Coding Based on Subspace Codes Outperforms Scalar Linear Network Coding}
\author{\IEEEauthorblockN{Tuvi Etzion and Antonia Wachter-Zeh}
\IEEEauthorblockA{Department of Computer Science\\
Technion---Israel Institute of Technology, Haifa, Israel\\
\emph{\{etzion, antonia\}@cs.technion.ac.il}
}}

\maketitle
\begin{abstract}
This paper considers vector network coding based on rank-metric codes and subspace codes.
Our main result is that vector network coding can significantly reduce the required field size
compared to scalar linear network coding in the same multicast network.
The achieved gap between the field size of scalar and vector network coding is in $q^{(h-2)t^2/h + o(t)}$
for any $q \geq 2$ and any even $h \geq 4$, where $t$ denotes 
the dimension of the vector solution and $h$ the number of messages.
If $h \geq 5$ is odd, then the
achieved gap of the field size between the scalar network coding solution
and the vector network coding solution is $q^{(h-3)t^2/(h-1) + o(t)}$.
Previously, only a gap of constant size had been shown.
This implies also the same gap between the field size in linear and non-linear scalar network coding
for multicast networks. 
The results are obtained by considering
several multicast networks which are variations of the well-known combination network.
\end{abstract}
\begin{keywords}
multicast networks, vector network coding, field size, combination network, rank-metric codes, subspace codes.
\end{keywords}

\section{Introduction}\label{sec:intro}
Network coding has been attracting increasing attention in the last fifteen years. The trigger for this interest was Ahlswede \emph{et al.}'s fundamental paper~\cite{Ahlswede_NetworkInformationFlow_2000} which revealed that network coding increases the throughput compared to simple routing.
An up-to-date survey on network coding for multicast networks can be found in~\cite{Fragouli-Sojanin-NetworkCodingMulticast_2015}.
In~\cite{KoetterMedard-AlgebraicApproachNetworkCoding_Journal}, K\"{o}tter and M\'{e}dard provided
an algebraic formulation for the network coding problem: for a given network, find coding
coefficients (over a small field) for each edge, which are multiplied with the symbols received at
the starting node of the edge, such that each receiver can recover all its requested information from its received symbols.
Such an assignment is called a \emph{solution} for the network.
If the coding coefficients are scalars,
it is called a \emph{scalar linear solution}.
Ebrahimi and Fragouli~\cite{EbrahimiFragouli-AlgebraicAlgosVectorNetworkCoding}
have extended this algebraic approach to vector network coding. Here, the received packets
are vectors and the coding coefficients are matrices. A set of coding matrices such that all receivers
can recover their requested information, is called a \emph{vector solution}.
In the sequel, we will consider only scalar linear network coding and vector linear network coding for multicast networks.

The \emph{field size} of the solution is an important parameter that directly influences
the complexity of the calculations at the network nodes.
Jaggi \emph{et al.}~\cite{JaggiSanders-PolyTimeAlgoMulticastNetworkCode}
have shown a deterministic algorithm for finding a network code (for multicast networks) of field size in the order of the number of receivers.
In general, finding the minimum required field size of a network code for a certain
multicast network is NP-complete~\cite{LehmanLehman-ComplexityNetworkInfoFlow}.

Since \emph{vector network coding} offers more freedom in choosing the coding
coefficients than scalar linear coding, a smaller field size might be achievable~\cite{DoughertyFreilingZeger-NetworksMatroidsNonShannon_2007}.
To our knowledge, Sun \emph{et al.}'s work~\cite{SunYangLongLi-MulticastNetworksVectorLinearCoding}
is the only one which presents explicit multicast networks where vector
network coding reduces the field size compared to scalar network coding.

This paper considers multicast networks, in particular a widely studied network,
the combination network, and several variations of it. We analyze the scalar and vector solutions of these networks.
The proposed vector solutions are based on rank-metric codes and subspace codes.
The main result of our paper is that for several of the analyzed networks,
our vector solutions significantly reduce the required \emph{field size}.
In these networks, the scalar solution requires a field size of $q^{(h-2)t^2/h+ o(t)}$, while we
provide a vector solution of field size $q$ and dimension~$t$, where the number of messages is an even number $h \geq 4$.
Therefore, the achieved gap between the scalar and the vector field size is 
$q^{(h-2)t^2/h + o(t)}$.
Throughout this paper, whenever we refer to such a \emph{gap}, we mean the difference between the \emph{smallest} field size for which a scalar linear network coding solution exists and the \emph{smallest} field size for which a vector network coding solution exists.
Similar results are given for an odd number of messages.
This improves upon~\cite{SunYangLongLi-MulticastNetworksVectorLinearCoding}, where only a constant gap,
which might be very large, was shown.
Further, the network of~\cite{SunYangLongLi-MulticastNetworksVectorLinearCoding} has a large
number of messages whereas our results are based on small and simple networks and hold for any number of messages
greater than two.
Finally, in the framework in~\cite{EbrahimiFragouli-AlgebraicAlgosVectorNetworkCoding},
the coding matrices for vector network coding have to be commutative, while in our solutions they
are not necessarily commutative.

This paper is structured as follows.
Section~\ref{sec:prelim} provides notations and definitions. Section~\ref{sec:networks}
defines the combination network and in Section~\ref{sec:vector-coding-combination},
we present a vector solution for the combination network.
In Section~\ref{sec:comb-extra-link-removed-receivers}, we present scalar and vector solutions
to modified combination networks with additional links.
For these networks, the required field size is significantly reduced and the gaps in the field sizes are derived.
In Section~\ref{sec:subspaces}, we show that the constructions which are based
on rank-metric codes can be seen as constructions based on subspace codes.
Moreover, using subspace codes, for additional
networks, the alphabet size can be reduced by using vector coding instead of scalar coding.
Concluding remarks and open problems are given in Section~\ref{sec:conclusion}.

Due to space limitations some proofs are only sketched and some are omitted
and can be found in the arxiv version~\cite{EtzionWachterzeh-Networkcoding},
where additional related material will be given.
Also, the most definitions of network coding are omitted.

\section{Preliminaries}\label{sec:prelim}
\subsection{Finite Fields and Subspaces}
Let $q$ be a power of a prime and let $\Fq$ denote the finite field of order $q$ and $\Fqm$ its extension field of order~$q^m$.
We use $\Fq^{m \times n}$ for the set of all $m\times n$ matrices over $\Fq$. 
Let $\mathbf I_{s}$ denote the $s \times s$ identity matrix and
$\0_s$ the $s\times s$ all-zero matrix. 

The triple $\codelinearArb{n,k,d}_{q}$ denotes a linear code over $\Fq$ of length~$n$, dimension $k$, and minimum Hamming distance~$d$.

Let $\langle \Mat{A} \rangle$ denote the space spanned by the rows of a matrix~$\Mat{A}$.
The \emph{Grassmannian} of dimension $r$, denoted by $\Grassm{n,r}$, is the set of all subspaces of $\Fq^n$ of dimension $r \leq n$.
The cardinality of $\Grassm{n,r}$ is the $q$-binomial coefficient:
\begin{equation*}
\big|\Grassm{n,r}\big|=\quadbinom{n}{r} \triangleq \prod\limits_{i=0}^{r-1} \frac{q^n-q^i}{q^r-q^i},
\end{equation*}
where
$q^{r(n-r)}\leq \quadbinom{n}{r} < 4 q^{r(n-r)}$.
For two subspaces $\myspace{U},\myspace{V}$, let $\myspace{U}+\myspace{V}$ denote the smallest subspace containing the union of $\myspace{U}$ and $\myspace{V}$.
The \emph{subspace distance} between $\myspace{U}$ and $\myspace{V}$ is defined by
$\Subspacedist{\myspace{U},\myspace{V}}
\triangleq2 \dim(\myspace{U}+\myspace{V})-\dim(\myspace{U})-\dim(\myspace{V})$.

\subsection{Rank-Metric Codes}
Let $\rk(\A)$ be the rank of $\A\in \Fq^{m \times n}$.
The \emph{rank distance} between $\A , \B \in \Fq^{m\times n}$ is defined by
$d_{\fontmetric{R}}(\A,\B)\triangleq  \rk(\A-\B)$.
A linear $\codelinearRank{m \times n,k,\delta}$ {rank-metric code} \mycode{C} is a $k$-dimensional linear subspace of $\Fq^{m \times n}$. It consists of $q^k$ matrices of size $m \times n$ over $\Fq$ with minimum rank distance
$\delta \triangleq
\min_{\substack{{\A} \in \mycode{C}, \A \neq \0}}
\big\lbrace \rk(\A) \big\rbrace$.
The Singleton-like upper bound for rank-metric
codes~\cite{Delsarte_1978,Gabidulin_TheoryOfCodes_1985,Roth_RankCodes_1991}
implies that for any $\codelinearRank{m \times n,k,\delta}$ code, we have $k \leq \max\{m,n\}(\min\{n,m\}-\delta+1)$.
Codes which attain this bound with equality are known
for all feasible parameters~\cite{Delsarte_1978,Gabidulin_TheoryOfCodes_1985,Roth_RankCodes_1991}.
They are called \emph{maximum rank distance} (MRD) codes and denoted by $\MRDlinq{m\times n, \delta}$.

A \emph{companion matrix} of a polynomial $p(x)$ is a $\deg p\times \deg p$ matrix consisting
of ones in the main sub-diagonal, the additive inverses of the coefficients of $p$ in the rightmost column, and zero elsewhere.
Let $\Mat{C}$ be the companion matrix of a primitive polynomial of degree $t$ over $\F_q$.
The set of matrices $\mycode{D}_t =\{ \0_t , \I_t , \Mat{C} ,\Mat{C}^2 , \ldots , \Mat{C}^{q^t-2} \}$
forms an $\MRDlinq{t\times t, t}$ code of~$q^t$ \emph{commutative} matrices (see also~\cite{Lusina2003Maximum}) which is isomorphic to $\F_{q^t}$.
These matrices are very useful when we design a network code for the combination network. Moreover, to prove that any network (multicast or non-multicast)
has a vector network code of dimension $t$ over $\F_q$ if the scalar solution is
over $\F_{q^t}$, we can use the set of matrices~$\mycode{D}_t$ as follows. Instead of the field elements in the scalar
network code, their vector representation with respect to a primitive element $\alpha$ in $\F_{q^t}$ is
used; instead of a coefficient $\alpha^s$ in the scalar solution, the matrix $\Mat{C}^s$ is used in the
vector solution, and instead of a zero coefficient the all-zero matrix is used. The matrices of $\mycode{D}_t$
are also very useful in encoding and decoding used in the network. Instead of computing
in the field $\F_{q^t}$, we can use the related matrices of the code to obtain the vector solution
and translate it to the scalar solution only at the receivers.

\section{The Combination Network}\label{sec:networks}
\begin{figure}[htb, scale = 1]
\centering
\tikzsetnextfilename{2-butterflies}
\begin{tikzpicture}[scale = 1]
	\node[mycircle,label=right:{$\vec{x}_1,\dots,\vec{x}_h$}] (sourcex) {} ; 
	\node[mycircle,below left=30pt and 40pt of sourcex] (middle0) {} ;
	\node[mycircle,below left=30pt and 20pt of sourcex] (middle1) {} ;
	\node[mycircle,below left=30pt and 0pt of sourcex] (middle2) {} ;
	\node[rectangle,right = 10pt of middle2](text1){$\dots$};
	\node[mycircle,below left=30pt and -60pt of sourcex] (middle3) {} ;
	\node[mycircle,below left=30pt and -80pt of sourcex] (middle4) {} ;
	\node[rectangle,right = 5pt of middle4](text2){$r$ nodes};

	\draw[black,->,very thick] (sourcex.south) -- (middle0.north);
	\draw[black,->,very thick] (sourcex.south) -- (middle1.north);
	\draw[black,->,very thick] (sourcex.south) -- (middle2.north);
	\draw[black,->,very thick] (sourcex.south) -- (middle3.north);
	\draw[black,->,very thick] (sourcex.south) -- (middle4.north);

	\draw [-, very thick,black] (2.4,-2.3) arc (160:70:10pt);
	\draw [-, very thick,black] (-1.8,-2.4) arc (140:30:13pt);
	\draw [-, very thick,black] (-2.5,-2.3) arc (130:20:10pt);
	
	\node[mycircle,below left=30pt and 10pt of middle0] (rec1) {} ;
	\node[mycircle,below left=30pt and -20pt of middle0] (rec2) {} ;
	\node[mycircle,below left=30pt and -150pt of middle0] (reclast) {} ; 
	
	\draw[black,->,very thick] (middle0.south) -- (rec1.north);
	\draw[black,->,very thick] (middle1.south) -- (rec1.north);
	\draw[black,->,very thick] (middle0.south) -- (rec2.north);
	\draw[black,->,very thick] (middle2.south) -- (rec2.north);
	\draw[black,->,very thick] (middle3.south) -- (reclast.north);
	\draw[black,->,very thick] (middle4.south) -- (reclast.north);
	\node[rectangle,below right = 7pt and 7pt of middle4](text2){$s$ edges};
	
\end{tikzpicture}
\caption{The $\mathcal{N}_{h,r,s}$-combination network. }
\label{fig:comb-net}
\end{figure}
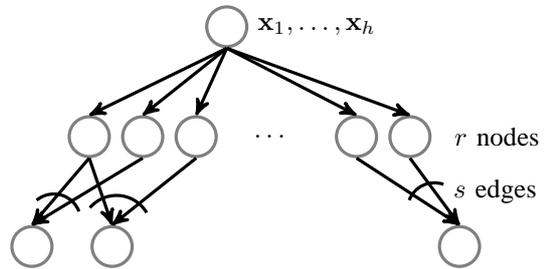

The $\mathcal{N}_{h,r,s}$-combination network is shown in Fig.~\ref{fig:comb-net} (see also~\cite{RiisAhlswede-ProblemsNetworkCodingECC}).
The network has three layers: in the first layer there is a source with $h$ messages.
The source transmits~$r$ new messages to the $r$ nodes of the middle layer, one
message to each node. Any $s$ nodes in the middle layer are
connected to a receiver, and each of the $\binom{r}{s}$ receivers demands all the original $h$ messages.
For vector coding, the messages $\vec{x}_1,\dots, \vec{x}_h$ are vectors of length $t$;
for scalar coding, the messages are scalars, denoted by $x_1, \dots, x_h$.

The $\mathcal{N}_{h,r,h}$-combination network has a scalar solution
of field size $q_s$ if and only if an $\codelinearArb{r,h,d=r-h+1}_{q_s}$ MDS code exists~\cite{RiisAhlswede-ProblemsNetworkCodingECC}.
Thus, $q_s \geq r-1$ if $q_s$ is odd and $q_s \geq r-2$ if $q_s$ is a power of 2 and  $h \in \{3,q_s-1\}$ are
sufficient~\cite[p.~328]{MacWilliamsSloane_TheTheoryOfErrorCorrecting_1988}.
The symbols which are transmitted from the source to each of the nodes in the middle layer form
together a codeword of the MDS code (encoded from the~$h$ message symbols).
Each receiver obtains~$h$ symbols from $h$ nodes of the middle layer.
Each receiver can correct $r-h$ erasures and hence it can reconstruct the~$h$ message symbols.

\section{Vector Coding in the Combination Network}
\label{sec:vector-coding-combination}
This section presents a vector solution based on MRD codes for the $\mathcal{N}_{h,r,h}$-combination network.
The case $h=2$ was implicitly already solved in a similar way in~\cite{SunYangLongLi-MulticastNetworksVectorLinearCoding}.

\subsection{Vector Linear Solution}
\begin{theorem}\label{thm:q-vand-block-matrix}
Let $\mycode{D}_t$ 
be the $\MRDlinq{t \times t,t}$ code defined by the companion matrix $\C$.
Let $\Mat{C}_i$, $i=1,\dots,h$, be distinct codewords of~$\mycode{D}_t$.
Define the following $ht \times ht$ block matrix:
\begin{equation*}
\Mat{M} =
\begin{pmatrix}
\I_t & \Mat{C}_1 & \Mat{C}_1^2 & \dots & \Mat{C}_1^{{h-1}}\\
\I_t & \Mat{C}_2 & \Mat{C}_2^2 & \dots & \Mat{C}_2^{{h-1}}\\
\vdots & \vdots& \vdots & \ddots & \vdots\\
\I_t & \Mat{C}_{h} & \Mat{C}_{h}^2 & \dots & \Mat{C}_{h}^{{h-1}}\\
\end{pmatrix}.
\end{equation*}
Then, any  $\ell t \times \ell t$ submatrix consisting of $\ell^2$ blocks of any $\ell t$
\textbf{consecutive columns} and any $\ell t$ \textbf{consecutive rows} has full rank $\ell t$, for any $\ell=1,\dots,h$.
\end{theorem}

\begin{construction}
\label{constr:comb-network-gen}
Let $\mycode{D}_t =\{\Mat{C}_1, \Mat{C}_2,\dots, \Mat{C}_{q^t}\}$ be the $\MRDlinq{t \times t,t}$
code defined by the companion matrix~$\C$ and let $r \leq q^t+1$. Consider the $\mathcal{N}_{h,r,h}$-combination
network with message vectors $\vec{x}_1, \dots, \vec{x}_h$.
One node from the middle layer receives and transmits $\vec{y}_{r} = \vec{x}_h$ and
the other $r-1$ nodes of the middle layer receive and transmit
$\vec{y}_{i} =
\begin{pmatrix}
\I_t \ \C_i \ \C_i^2 \ \dots \ \C_i^{h-1}
\end{pmatrix}
\cdot
\begin{pmatrix}
\vec{x}_1 \
\vec{x}_2 \
\dots \
\vec{x}_h
\end{pmatrix}^T
\in \Fq^t$,
for $i=1,\dots,r-1$.
\end{construction}
The matrices $\I_t,\C_i, \C_i^2,\dots,\C_i^{h-1}$ are the coding coefficients of the incoming and outgoing edges of the middle layer nodes.
\begin{theorem}\label{thm:solution-combination}
Construction~\ref{constr:comb-network-gen} provides a vector linear solution of field size $q$ and dimension $t$ to the $\mathcal{N}_{h,q^t+1,h}$-combination network, i.e., $\vec{x}_1, \dots, \vec{x}_h$ can be reconstructed at all receivers.
\end{theorem}
\begin{IEEEproof}
Each receiver obtains
\begin{equation*}
\begin{pmatrix*}
\vec{y}_{i_1}\\
\vdots\\
\vec{y}_{i_{h-1}}\\
\vec{y}_{i_h}
\end{pmatrix*}
=
\begin{pmatrix}
\I_t & \Mat{C}_{i_1} & \Mat{C}_{i_1}^2 & \dots & \Mat{C}_{i_1}^{{h-1}}\\
\I_t & \Mat{C}_{i_2} & \Mat{C}_{i_2}^2 & \dots & \Mat{C}_{i_2}^{{h-1}}\\
\vdots & \vdots& \vdots & \ddots & \vdots\\
\I_t & \Mat{C}_{i_h} & \Mat{C}_{i_h}^2 & \dots & \Mat{C}_{i_h}^{{h-1}}\\
\end{pmatrix}.
\begin{pmatrix}
\vec{x}_1\\
\vec{x}_2\\
\vdots\\
\vec{x}_h
\end{pmatrix}
\end{equation*}
or
\begin{equation*}
\begin{pmatrix*}
\vec{y}_{i_1}\\
\vdots\\
\vec{y}_{i_{h-1}}\\
\vec{y}_1
\end{pmatrix*}
=
\begin{pmatrix}
\I_t & \Mat{C}_{i_1} & \Mat{C}_{i_1}^2 & \dots & \Mat{C}_{i_1}^{{h-1}}\\
\vdots & \vdots& \vdots & \ddots & \vdots\\
\I_t & \Mat{C}_{i_{h-1}} & \Mat{C}_{i_{h-1}}^2 & \dots & \Mat{C}_{i_{h-1}}^{{h-1}}\\
\0_t & \0_t & \0_t & \dots & \I_t
\end{pmatrix}.
\begin{pmatrix}
\vec{x}_1\\
\vec{x}_2\\
\vdots\\
\vec{x}_h
\end{pmatrix},
\end{equation*}
for some distinct $i_1,\dots,i_h \in \{2,\dots,r\}$.
Due to Theorem~\ref{thm:q-vand-block-matrix}, in both cases, the corresponding matrix has full rank and there is a unique solution for $(\vec{x}_1 \ \vec{x}_2 \ \dots \ \vec{x}_h)$.
\end{IEEEproof}


For the $\mathcal{N}_{3,q^t+2,3}$-combination network and when~$q^t$ is a power of two, we can use the matrices from Construction~\ref{constr:comb-network-gen} and additionally $(\0_t \ \I_t \ \0_t) \cdot (\vec{x}_1 \ \vec{x}_2 \ \vec{x}_3)^T$
to obtain a vector linear solution. All the corresponding matrices have full rank.

\subsection{Analysis}
\label{subsec:comparison-comb}
Due to the isomorphism of $\F_{q^t}$ and the code $\mycode{D}_t$, both solutions are equivalent. Implementing the scalar solution can actually be done by implementing the vector solution.
We can therefore construct a vector linear solution of size $q$ and dimension $t$ for the $\mathcal{N}_{h,q^t+1,h}$-combination
network, where equivalently a scalar solution from an MDS code exists for $q_s \geq q^t$.
The decoding complexity when implementing the vector solution is in the order of
$\OCompl{th \log^2 (t) \log^2 (h)}$ operations over $\Fq$ for each receiver.

\section{A Generalization of the Combination Network with Extra Links}
\label{sec:comb-extra-link-removed-receivers}
\subsection{Considered Network}
\label{subsec:comb-net-rate-2}
\begin{figure}[htb, scale = 1]
\centering
\tikzsetnextfilename{2-butterflies}

  
\begin{tikzpicture}[scale = 1]
	\node[mycircle,label=right:{$\vec{x}_1,\vec{x}_2,\vec{x}_3,\vec{x}_4$}] (sourcex) {} 
	edge[bend left=5, very thick,->] (middle0.north)
	edge[bend right=5, very thick,->] (middle0.north)
	edge[bend left=5, very thick,->] (middle1.north)
	edge[bend right=5, very thick,->] (middle1.north)
	edge[bend left=5, very thick,->] (middle2.north)
	edge[bend right=5, very thick,->] (middle2.north)
	edge[bend left=5, very thick,->] (middle3.north)
	edge[bend right=5, very thick,->] (middle3.north)
	edge[bend left=5, very thick,->] (middle4.north)
	edge[bend right=5, very thick,->] (middle4.north)
	
	edge[bend right=40, very thick,->] (rec1.west)
	edge[bend left=40, very thick,->] (rec2.east)
	edge[bend right=20, very thick,->] (reclast.west)
	; 
	
	\node[mycircle,below left=30pt and 40pt of sourcex] (middle0) {} 
	edge[bend left=5, very thick,->] (rec1.north)
	edge[bend right=5, very thick,->] (rec1.north)
	edge[bend left=5, very thick,->] (rec2.north)
	edge[bend right=5, very thick,->] (rec2.north)
	
	; 
	\node[mycircle,below left=30pt and 20pt of sourcex] (middle1) {} 
	edge[bend left=5, very thick,->] (rec1.north)
	edge[bend right=5, very thick,->] (rec1.north);
	\node[mycircle,below left=30pt and 0pt of sourcex] (middle2) {} 
	edge[bend left=5, very thick,->] (rec2.north)
	edge[bend right=5, very thick,->] (rec2.north);
	\node[rectangle,right = 10pt of middle2](text1){$\dots$}
	;
	\node[mycircle,below left=30pt and -60pt of sourcex] (middle3) {} 
	edge[bend left=5, very thick,->] (reclast.north)
	edge[bend right=5, very thick,->] (reclast.north);
	\node[mycircle,below left=30pt and -80pt of sourcex] (middle4) {}
	edge[bend left=5, very thick,->] (reclast.north)
	edge[bend right=5, very thick,->] (reclast.north) ;
	\node[rectangle,right = 5pt of middle4](text2){$r$ nodes};

	
	\node[mycircle,below left=30pt and 10pt of middle0] (rec1) {} ;
	\node[mycircle,below left=30pt and -20pt of middle0] (rec2) {} ;
	\node[mycircle,below left=30pt and -150pt of middle0] (reclast) {} ; 
	
	
\end{tikzpicture}
\caption{The $\mathcal{N}^*_{h,r,h}$-network, drawn for $h=4$, $\ell=2$. }
\label{fig:comb-net-rate2}
\end{figure}
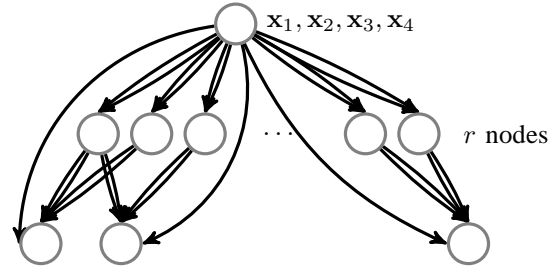

In this section, we modify the combination network.
We consider the $\mathcal{N}^*_{h,r,h}$-network, shown in Fig.~\ref{fig:comb-net-rate2}, first for $h=2 \ell=4$.
It has three layers, a source in the first layer and~$r$ nodes in the middle layer, with two links from the source to each node in the middle layer. There are $\binom{r}{2}$ receivers in the third layer, where any two nodes from the middle layer are connected to
a different receiver. If a node $U$ from the middle layer is connected to a receiver $R$, then there are two links from $U$ to $R$.
There is also a direct link from the source to each receiver.
The structure of this network differs from most networks in the literature since the
min-cut between the source and each receiver is $h+1$ (and not $h$) and there are parallel edges.
In Section~\ref{sec:conclusion}, we show how to transform this to an equivalent network with min-cut $h$
and without parallel edges.

\subsection{Scalar Linear Solution}
\begin{lemma}\label{lem:scalar-lin-rate-two}
There is a scalar linear solution of field size $q_s$ for the $\mathcal{N}^*_{4,r,4}$-network if and only if
$r \leq (q_s^2+1)(q_s^2+q_s+1)$.
\end{lemma}
\begin{IEEEproof}
Let $\Mat{B}$ be a $4 \times 2r $ matrix, divided into $r$ blocks of two columns,
with the property that any two blocks together have rank at least three.
From each one of the~$r$ nodes in the middle layer, transmit two symbols
(from one block) of $(x_1, x_2,x_3,x_4) \cdot \Mat{B}$ (these symbols were
transmitted to the node from the source). On each extra link,
transmit a symbol $p_i = \sum_{j=1}^{4} p_{ij} x_j$, for $i=1,\dots,\binom{r}{2}$,
which is chosen such that the corresponding $4 \times 4$-submatrix of $\Mat{B}$ with the
additional column $(p_{i1}, p_{i2}, p_{i3},  p_{i4})^T$ has full rank four.
There is a scalar solution over $\F_{q_s}$ if and only if such a matrix over $\F_{q_s}$ exists.

Define these blocks to be any $4\times 2$-matrix representations of all $2$-dimensional subspaces
of $\F_{q_s}^4$. Any two blocks together form a $4\times 4$ matrix of rank at least three (since any two such subspaces are distinct).

From every node in the middle layer, there are two links to the appropriate receivers.
Therefore, we associate each middle node with one block. The number of blocks is at most the
number of distinct 2-dimensional subspaces of $\F_{q_s}^4$, i.e. $r \leq \quadbinoms{4}{2}$
and therefore, a scalar solution exists if:
\begin{equation*}
r \leq \quadbinoms{4}{2} = (q_s^2+1)(q_s^2+q_s+1).
\end{equation*}

To prove the "only if", we show that there is no scheme that provides more blocks.
Assume, one block is a matrix of rank one. Then, all other blocks must have rank two and the space that they span has
to be disjoint to the block of rank one. Therefore, with this scheme there are $1 + \quadbinoms{3}{2} < \quadbinoms{4}{2}$ blocks.
Thus, to maximize $r$, all blocks should have rank two, and taking all
distinct $2$-dimensional subspaces yields the maximum number of blocks.
\end{IEEEproof}

\subsection{Vector Linear Solution}
\label{subsec:vec-2-rate-k=4}
\begin{construction}
\label{constr:mrd-notfull-rank-comb-2}
Let $\mycode{C}=\{\Mat{C}_1,\Mat{C}_2,\dots,\Mat{C}_{q^{2t^2 + 2t}}\}$ be an $\MRDlinq{2t \times 2t,t}$ code and let $r \leq q^{2t^2 + 2t}$.
Consider the $\mathcal{N}^*_{4,r,4}$-network with message vectors $\vec{x}_1, \vec{x}_2, \vec{x}_3, \vec{x}_4 \in \Fq^t$.
The $i$-th middle node receives and transmits:\\[-1.8ex]
\begin{equation*}
\begin{pmatrix}
\vec{y}_{i_1} \\
\vec{y}_{i_2}
\end{pmatrix}
=
\begin{pmatrix}
\I_{2t} & \C_i
\end{pmatrix}
\cdot
\begin{pmatrix}
\vec{x}_1\\
\vec{x}_2\\
\vec{x}_3\\
\vec{x}_4
\end{pmatrix}
\in \Fq^{2t},
\quad
i=1,\dots,r.
\end{equation*}
The extra link from the source which ends in the same receiver as the links from two distinct nodes $i,j \in \{ 1,2,\ldots , r\}$,
of the middle layer transmits the vector $\vec{z}_{ij} = \Mat{P}_{ij} \cdot \begin{pmatrix}
\vec{x}_1,
\vec{x}_2,
\vec{x}_3,
\vec{x}_4
\end{pmatrix}^T\in \Fq^t$,
where the $t \times 4t$ matrix $\vec{P}_{ij}$ is chosen such that
\begin{equation*}
\rk\begin{pmatrix}
\I_{2t} & \C_i \\
\I_{2t} & \C_j \\
&\hspace{-4ex} \Mat{P}_{ij}
\end{pmatrix} = 4t.
\end{equation*}
\end{construction}
Clearly, $\rk\left(\begin{smallmatrix}
\I_{2t} & \C_i \\
\I_{2t} & \C_j \\
\end{smallmatrix}\right) \geq 3t$, and hence the $t$ rows of $\Mat{P}_{ij}$ can be chosen such that the overall rank is $4t$.
\begin{theorem}
\label{thm:solution-extra-rate2}
Construction~\ref{constr:mrd-notfull-rank-comb-2} provides a vector solution of field size $q$
and dimension $t$ to the $\mathcal{N}^*_{4,r,4}$-network for $ r\leq q^{2t(t+1)}$.
\end{theorem}
\begin{IEEEproof} 
On each receiver, we obtain
\begin{equation*}
\begin{pmatrix*}
\vec{y}_{i_1}\\
\vec{y}_{i_2}\\
\vec{y}_{j_1}\\
\vec{y}_{j_2}\\
{\vec{z}_{ij}}
\end{pmatrix*}
=
\begin{pmatrix}
\I_{2t} & \C_i\\
\I_{2t} & \C_j\\
& \hspace{-4ex}\Mat{P}_{ij}
\end{pmatrix}
\cdot
\begin{pmatrix}
\vec{x}_1\\
\vec{x}_2\\
\vec{x}_3\\
\vec{x}_4
\end{pmatrix}.
\end{equation*}
The choice of $\Mat{P}_{ij}$ guarantees that this linear system of equations
has a unique solution for $(\vec{x}_1, \vec{x}_2,\vec{x}_3,\vec{x}_4)$.
\end{IEEEproof}

\subsection{Comparison of the Solutions}
For the $\mathcal{N}^*_{4,r,4}$-network, we obtain a significant improvement in the field size for vector
coding compared to scalar coding. The field size of the vector coding solution is equivalent
to~$q^t$ while in scalar coding, $r \leq (q_s^2+1)(q_s^2+q_s+1)$.
Since $r$ can be chosen to be $q^{2t^2+2t}$, we have that the gap size is $q^{t^2/2 + o(t)}$.

\subsection{Arbitrary Number of Messages}
\label{sec:gen_k}
Let us shortly outline the case of $h=2 \ell$ messages, where $\ell \geq 2$.
The $\mathcal{N}^*_{h,r,h}$-network has three layers, a source in the first layer
and $r$ nodes in the middle layer. The source is connected with $\ell$ parallel edges to each node in the middle layer.
There are~$\binom{r}{2}$ receivers and a link from the source to each receiver. Each two nodes from the middle layer are
connected to exactly one receiver. If node $U$ from the middle layer is connected to receiver $R$,
then there are $\ell$ parallel edges from $U$ to $R$. Thus, each receiver gets $2\ell +1$ links in total; namely,
$2\ell$ links from two middle nodes and one link from the source.
The optimal scalar solution is obtained when it is considered
that each middle node is transmitting $\ell$ blocks, each one with $2 \ell$ symbols from the alphabet $q_s$.
In the optimal solution each of these $\ell$ blocks forms an $\ell$-dimensional subspace of~$\F_{q_s}^{2\ell}$ such
that two such $\ell$-dimensional subspaces intersect in a subspace of dimension at most one.
In other words, the subspace distance between two such sets is at least $2 \ell -2$.
The size of the largest set with such $\ell$-dimensional subspaces in $\F_{q_s}^{2\ell}$ is
of the order $q^{2\ell}$~\cite{Etzion_Storme_2015}. For the vector solution, we
can use an $\MRDlinq{\ell t \times \ell t,(\ell-1)t}$ code whose size is $q^{\ell t^2 +\ell t}$.
Thus, we have that the gap size is $q^{t^2/2 + o(t)}$, for any $h=2 \ell$, $\ell \geq 2$.

To improve these results,
we need another generalization of the $\mathcal{N}^*_{4,r,4}$-network.
The new network will be called the $\mathcal{N}^+_{k,r,k}$-network. It has three layers,
with a source carrying $h=2\ell$ messages in the first layer. In the second layer
there are $r$ nodes and in the third layer there are $\binom{r}{2}$ receivers.
The links between the source and the nodes of the second layer and between the
nodes of the second layer and the receivers are the same as in the $\mathcal{N}^*_{h,r,h}$-network.
The $\mathcal{N}^*_{h,r,h}$-network and the $\mathcal{N}^+_{h,r,h}$-network differ
in the number of links between the source and each receiver. While in the $\mathcal{N}^*_{h,r,h}$-network
there is exactly one link between the source and each receiver, in the $\mathcal{N}^+_{h,r,h}$-network
there are $\ell-1$ links from the source to each receiver. Note, that $\mathcal{N}^*_{4,r,4}=\mathcal{N}^+_{4,r,4}$.
The scalar solution and the vector solution are also very similar in this generalization
to the solution for $\ell =2$. The optimal scalar solution is obtained when we consider
that a node in the middle layer is transmitting $\ell$ blocks, each one with $2 \ell$ symbols from the alphabet $q_s$.
In the optimal solution each of these $\ell$ blocks forms an $\ell$-dimensional subspace of $\F_{q_s}^{2\ell}$ such
that two such $\ell$-dimensional subspaces intersect in at most an $(\ell-1)$-dimensional subspace.
In other words, the subspace distance between two such sets is at least $2$, i.e. all $\ell$-dimensional subspaces
of $\Grassm{2 \ell , \ell}$. The size of $\Grassm{2 \ell , \ell}$  is $\quadbinom{2 \ell}{\ell}$ which is
of the order $q^{\ell^2}$~\cite{Etzion_Storme_2015}. For the vector solution, we
can use an $\MRDlinq{\ell t \times \ell t,t}$ code whose size is $q^{\ell (\ell -1) t^2 +\ell t}$.
Thus, we have that the gap size is $q^{(\ell-1)t^2/\ell + o(t)}$.

For an odd number of messages $2\ell +1$, $\ell \geq 2$, we can use the modifications of the networks
$\mathcal{N}^*_{2\ell,r,2\ell}$ and $\mathcal{N}^+_{2\ell,r,\ell}$ with an additional
link from the source to each receiver to obtain similar results to the ones
with even number of messages. A network with $h=3$ messages is discussed in the next section.

\section{Vector Solutions Using Subspace Codes}
\label{sec:subspaces}
Our constructions from the previous sections are based on rank-metric codes,
but can be seen as a special case of a more general construction
based on subspace codes. In the sequel, we explain
the simple formulation of this construction, demonstrate how one
of our constructions can be improved by using subspace codes,
and present a general question on subspace codes which is derived from our discussion.
Finally, we show a multicast network with three messages in which vector network
coding outperforms scalar network coding, where the key is to use special
classes of subspace codes.

The formulation with subspaces can be derived by noticing that the rows of the matrix
$
[\I_t ~\Mat{C}],
$
where $\Mat{C}$ is a $t \times n$ matrix, is a basis of a subspace of dimension $t$ in $\F_q^{t+n}$ and the set of all such
matrices in the network code forms a code in $\Grassm{t+n,t}$. For various networks and constructions,
we have to understand what kind of code is required for each network.

For example, Construction~\ref{constr:mrd-notfull-rank-comb-2} in Section~\ref{subsec:vec-2-rate-k=4} can be improved
by using a code in $\Grassm{4t,2t}$ with minimum subspace distance $2t$.
A basis for a codeword is a $2t \times 4t$ matrix and the matrices which
form the basis for the codewords can replace
the $2t \times 4t$ matrices of the form $[\I_{2t} ~ \Mat{C}_i]$ in Construction~\ref{constr:mrd-notfull-rank-comb-2}.
Such a code will enable us to use more nodes in the middle layer
of the network. Constructions of large codes for this purpose can be found
for example in~\cite{Etzion2009ErrorCorrecting}. However, the improvement is not large since
asymptotically the code obtained from an MRD code which was used in
Construction~\ref{constr:mrd-notfull-rank-comb-2} is optimal and can be improved by
at most a factor of four~\cite{Etzion_Storme_2015}.

Also for the other constructions, e.g., the generalizations in Section~\ref{sec:gen_k},
subspace codes can be used. For these constructions and other variations, the required
large subspace code is described as follows.
For a given $\rho$, $0 \leq \rho \leq \ell-2$, find a large code
in $\Grassm{ {\ell}t ,t}$ such that
the linear span of the rows of any $\ell$ codewords is a subspace
whose dimension is at least $(\ell - \rho )t$. Such a code can be used when
$\rho$ links connect the source with each receiver.
More generalizations will be discussed in the full version of this paper.

One example of such construction which requires a new type of subspace codes
is a multicast network with three messages in which vector network coding
outperforms scalar network coding. The network is a simple modification of the
$\mathcal{N}_{3,r,3}$-combination network. The new network $\mathcal{\widetilde{N}}_{3,r,3}$
consists of $\mathcal{N}_{3,r,3}$ with an additional link from the source to each receiver.
For scalar network coding, each edge carries three coefficients
which can be viewed as a one-dimensional subspace of $\F_{q_s}^3$. At most two edges from the
three edges, originating in the middle layer and
ending in the same receiver, can carry the same one-dimensional subspace.
Hence, $r \leq 2(q_s^2+q_s+1)$.
We demonstrate the advantage of vector network coding on scalar network coding
on a specific example. Assume $q=4=2^2$, i.e., $r \leq 42$, and consider now vector network coding,
where the messages are binary vectors of length $t=2$. Hence, the edges will carry
2-dimensional subspaces of $\F_2^6$. Vector network coding will outperform
scalar network coding if we will find more than 42 2-dimensional subspaces
of $\F_2^6$ such that any three 2-dimensional subspace will span at least a 4-dimensional
subspace, so they will be completed by the extra link from the source to the receiver.
Such a code with many more than 42 subspaces can be found using certain spreads.
This method can be generalized for other parameters and will be discussed in the full version of the paper.

\section{Concluding Remarks and Future Work}
\label{sec:conclusion}

We have shown that vector network coding outperforms scalar linear network coding
in the alphabet size for several variations of the combination network.
The key is the use of subspace codes and in particular subspace codes derived from rank-metric codes.

It should be remarked that the min-cut in our modified combination networks is larger than the number of messages.
This can be fixed as follows: replace the $i$-th receiver $R_i$ by a node $T_i$
from which there are $h$ links to $h$ vertices $P_{ij}$, $1 \leq j \leq h$.
From $P_{ij}$, $1 \leq j \leq h$, there is a link to a new receiver $R_i^\prime$. The
new network is solvable with the same alphabet as the old network, and the min-cut in the new network is~$h$.
Similarly we can avoid parallel links in the network. Assume there are $\ell$ parallel links
from vertex $U$ to vertex $V$. We can remove these links, add $\ell$ vertices $W_1,W_2,\ldots ,W_\ell$,
such that there exists a link from $U$ to $W_i$, $1 \leq i \leq \ell$, and there exists a link
from each vertex $W_i$, $1 \leq i \leq \ell$, to $V$. Again, the new network will be solved with the same
alphabet as the old network. In our specific networks it can be done more efficiently
by replacing each node in the middle layer by $\ell$ nodes.

Clearly, a vector network code can be translated to a non-linear scalar network
code. Therefore, our results also imply a gap of size $q^{(h-2)t^2/h + o(t)}$
(for even $h\geq 4$)
between the field size in linear and non-linear scalar network coding
for multicast networks. 

Some open questions for future research are briefly outlined as follows:
\begin{itemize}
\item Design a network with two messages in which vector network coding outperforms
scalar network coding in the alphabet size or show that such a network does not exist.

\item For each number of messages $h$, find the largest possible gap in the alphabet size
between the solutions of scalar linear network coding and vector network coding.

\item Is there a network with $h$ messages in which exactly $h$ edge disjoint paths
are used (for network coding) from the source
to each receiver, and on which vector coding
outperforms scalar linear network coding w.r.t the field size?
Note that our constructions use more than $h$ paths.

\item Construct subspace codes with the required properties outlined
in Section~\ref{sec:subspaces}.

\end{itemize}

Finally, we have considered several more related networks and their description together
with a comparison of vector network coding and scalar network coding. These networks
will appear in the full version of this paper~\cite{EtzionWachterzeh-Networkcoding}.

\section*{Acknowledgements}
T. Etzion's work was supported in part by the Israeli Science Foundation (ISF), Jerusalem, Israel, under Grant 10/12.

A. Wachter-Zeh's work was supported by the European Union’s Horizon 2020 research
and innovation programme under the Marie Sklodowska-Curie grant agreement No. 655109.

The authors thank the three anonymous reviewers whose
careful reading and insightful comments have contributed to the presentation
of this paper.

\bibliographystyle{IEEEtranS}
\bibliography{antoniawachter}

\end{document}